\documentclass[aps,prl,reprint,groupedaddress,superscriptaddress,preprintnumbers]{revtex4-2}

\usepackage{subfigure}
\usepackage{graphicx,natbib}
\usepackage{bm}
\usepackage{color}
\usepackage{amssymb}
\usepackage{hyperref}
\usepackage{mathrsfs}

\newcommand{\AAA}{\bm{A}}
\newcommand{\BB}{\bm{B}}
\newcommand{\UU}{\bm{U}}

\newcommand{\SSSS}{\mbox{\boldmath ${\sf S}$} {}}

\newcommand{\meanrho}{\overline{\rho}}
\def\cs{c_{\rm s}}
\newcommand{\nab}{\bm{\nabla}}
\def\Rm{{\rm Re_\mathrm{M}}}

\newcommand{\meanBB}{\overline{\mbox{\boldmath $B$}}{}}{}
\newcommand{\meanv}{\overline{v}}

\newcommand{\meanEMF}{\overline{\mbox{\boldmath ${\cal E}$}}{}}{}
\newcommand{\meanmu}{\overline{\mu}}

\def\jcap{{J.\ Cosmol.\ Astropart.\ Phys.}}
\def\apjl{{Astrophys.\ J.\ Lett.}}

\newcommand{\bra}[1]{\langle #1\rangle}
\def\cs{c_{\rm s}}
\def\vA{v_{\rm A}}

\def\kB{k_{\rm B}}

\newcommand{\alphaem}{\ensuremath{\alpha_{\rm em}}}

\begin{document}

\title{Production of a Chiral Magnetic Anomaly with Emerging Turbulence \\ and Mean-Field Dynamo Action}
\preprint{NORDITA-2021-066}

\author{Jennifer~Schober}
\email{jennifer.schober@epfl.ch}
\affiliation{Laboratoire d'Astrophysique, EPFL, CH-1290 Sauverny, Switzerland}

\author{Igor~Rogachevskii}
\affiliation{Department of Mechanical Engineering, Ben-Gurion University of the Negev, P.O. Box 653, Beer-Sheva 84105, Israel}
\affiliation{Nordita, KTH Royal Institute of Technology and Stockholm University, 10691 Stockholm, Sweden}

\author{Axel~Brandenburg}
\affiliation{Nordita, KTH Royal Institute of Technology and Stockholm University, 10691 Stockholm, Sweden}
\affiliation{The Oskar Klein Centre, Department of Astronomy, Stockholm University, AlbaNova, SE-10691 Stockholm, Sweden}
\affiliation{School of Natural Sciences and Medicine, Ilia State University, 0194 Tbilisi, Georgia}
\affiliation{McWilliams Center for Cosmology and Department of Physics, Carnegie Mellon University, Pittsburgh, Pennsylvania 15213, USA}

\date{\today}

\begin{abstract}
In relativistic magnetized plasmas, asymmetry in the number densities of left- and right-handed fermions, i.e., a nonzero chiral chemical potential $\mu_5$, 
leads to an electric current along the magnetic field.
This causes a chiral dynamo instability for a uniform $\mu_5$, but our simulations reveal a dynamo 
even for fluctuating $\mu_5$ with zero mean.
It produces magnetically dominated turbulence and generates
mean magnetic fields via the magnetic $\alpha$ effect.
Eventually, a universal scale-invariant $k^{-1}$ spectrum of $\mu_5$ and 
a $k^{-3}$ magnetic spectrum are formed independently of the initial condition. 
\end{abstract}

\maketitle

\noindent

The chiral magnetic effect (CME) is a macroscopic quantum phenomenon.
It leads to an electric current along
the magnetic field due to an imbalance between
oppositely handed electrically charged fermions \citep{Vilenkin:80a}.
This is a direct consequence of the coupling of fermionic chirality 
and the topology of magnetic field lines characterized by magnetic 
helicity \citep{KH14,K16}.
Chiral asymmetry is quantified by
the chiral chemical potential $\mu_5 \equiv \mu_\mathrm{L} -\mu_\mathrm{R}$, 
which is nonzero in regions where the chemical potentials
of left- $(\mu_\mathrm{L})$ and right-handed 
$(\mu_\mathrm{R})$ fermions differ.
It has been shown \citep{BFR12} that $\mu_5$ can survive down to energies 
of $\approx 10~\mathrm{MeV}$ and thereby the CME can potentially affect 
leptogenesis during the QCD phase transition \citep{SchwarzStuke2009} and produce
gravitational waves in the early Universe \citep{BrandenburgHeEtal2021}.

The dynamics of chiral fluids has been studied in various approaches 
\citep{BFR12,FigueroaEtAl2019,MaceEtAl2020,StephanovYin2012,ChenEtAl2015,GSV16,YamamotoYang2020}, 
including an effective description called
chiral magnetohydrodynamics (MHD) 
\citep{GI13,REtAl17,ZB18,HattoriEtAl2019}.
A significant difference to classical MHD is that the CME can induce a dynamo instability in 
the magnetic field on small length scales \citep{JS97}.
Unlike classical MHD dynamos, chiral dynamos
can occur without an initial velocity field
and self-consistently produce 
turbulence through the Lorentz force. 
This can activate a chiral mean-field dynamo 
\citep{REtAl17,Schober2017,SBR19,SBR20}.

The possibility of efficient magnetic field 
amplification through the CME has relevance for
the early Universe. 
In particular, the transport of magnetic energy to large 
length scales via a chiral inverse cascade 
\citep{BFR12,HKY15,GRS16,BSRKBFRK17}
and the chiral mean-field dynamo,
strongly increases the chance of primordial magnetic fields \citep{S16,Vachaspati2021}
to survive until present day. 
Thereby, observational constraints on magnetic fields in 
cosmic voids \citep{NV10} may open up a unique window 
into the fundamental physics of the early Universe.
Beyond cosmology, chiral MHD has also relevance to neutron stars
\citep{DS18,MKT18,Yamamoto:2015gzz,SiglLeite2016,DvornikovEtAl2020},
quark-gluon plasmas in heavy-ion collisions \citep{KH14,K16,HKY17}, 
and quantum materials \citep{GalitskiEtAl2018}.

In all previous chiral dynamo studies, 
a uniform initial $\mu_5$ has been considered
\citep{REtAl17,Schober2017,SBR19,SBR20}.
However, a uniform $\mu_5$ requires
special generation mechanisms. 
Therefore, we consider in this Letter a more general 
and universal situation with 
initial fluctuations of the chiral chemical potential,
but zero mean.

For the analysis, 
we normalize $\mu_5$ by $4 \alphaem / (\hbar c)$
such that it has the dimension of inverse length,
where $\alphaem$ is the fine structure constant,
$c$ is the speed of light, and $\hbar$ is the reduced Planck constant.
The strength of the coupling of the electromagnetic field
to $\mu_5$ is characterized by the chiral feedback parameter
$\lambda$ which, for hot plasmas, is given by
$\lambda=3 \hbar c (8 \alphaem)^2/(\kB T)^2$,
where $T$ is the temperature and $\kB$ is the Boltzmann constant.
We consider the following set of chiral MHD equations \citep{REtAl17}:
\begin{eqnarray}
  \frac{\partial \BB}{\partial t} &=& \nab   \times   \left[{\UU}  \times   {\BB}
  - \eta \, \left(\nab   \times   {\BB}
  - \mu_5 {\BB}  \right) \right] ,
\label{ind-DNS}\\
  \rho{D \UU \over D t}&=& (\nab   \times   {\BB})  \times   \BB-\nab  p + \nab  {\bm \cdot} (2\nu \rho \SSSS) ,
\label{UU-DNS}\\
  \frac{D \rho}{D t} &=& - \rho \, \nab  \cdot \UU
\label{rho-DNS} , \\
  \frac{D \mu_5}{D t} &=& \mathscr{D}_5(\mu_5)
  + \lambda \, \eta \, \left[{\BB} {\bm \cdot} (\nab   \times   {\BB}) - \mu_5 {\BB}^2 \right], 
  \label{mu5-DNS}
\label{mu-DNS}
\end{eqnarray}
where the magnetic field $\BB$ is normalized such that the magnetic energy 
density is $\BB^2/2$, and
$D/D t = \partial/\partial t + \UU \cdot \nab$ with
$\UU$ being the velocity field.
Further, $\eta$ is the microscopic magnetic diffusivity,
$p$ is the fluid pressure,
${\sf S}_{ij}=(U_{i,j}+U_{j,i})/2 -
\delta_{ij} ({\bm \nabla}{\bm \cdot} \UU)/3$
are the components of the trace-free strain tensor $\SSSS$ (commas denote
partial spatial derivatives) and
$\nu$ is the kinematic viscosity.
We adopt an isothermal equation of state, $p=\rho\cs^2$,
with $\cs$ being the sound speed.
Equations~(\ref{ind-DNS})--(\ref{mu-DNS}) imply 
that total chirality $\chi_{\rm tot} \equiv
\langle \mathcal{H}\rangle + 2 \langle \mu_5 \rangle /\lambda$
is conserved, where angle brackets denote volume averaging.
Here, $\langle\mathcal{H}\rangle \equiv \langle \AAA \cdot \BB\rangle$ is the magnetic helicity with the vector potential $\AAA$ and $\BB = \nabla \times \AAA$.

At the initial time $t_0$, we assume $\langle \mu_5 \rangle(t_0)=0$, but nonzero fluctuations, $\mu'_5$, i.e.,
$\langle {\mu'_5}^2 \rangle(t_0) \not=0$.
Initially, small fluctuations of $\BB$ 
with zero mean are present, while the velocity field vanishes.
The fluctuations $\mu'_5$ result in an exponential growth 
of magnetic fluctuations due to the chiral dynamo.
This instability is caused by 
the term $\nab\times(v_5 \BB)$ in Eq.~(\ref{ind-DNS}) with
$v_5=\eta \mu_5$
and has a growth rate $\gamma(k) = |v_5| k - \eta k^2$,
with $k$ being 
the wave number.
This instability is referred to 
as the chiral dynamo \citep{JS97}
and occurs when $|v_5| > \eta k$.
Its maximum growth rate is
$\gamma_5 = v_5^2/4 \eta$
and is attained at $k_5 = |\mu_5|/2$.
We note that, while the $\nab\times(v_5 \BB)$ term in Eq.~(\ref{ind-DNS}) 
is formally similar 
to the kinetic $\alpha$ effect 
in classical mean-field MHD \citep{REtAl17}, the velocity $v_5$ is not produced by helical turbulence, but rather by the CME.
During the chiral dynamo phase, 
magnetic fluctuations produce velocity fluctuations via 
the Lorentz force $(\nab  \times {\BB}) \times \BB$.

Since the initial mean chiral chemical potential is zero, 
and the initial small-scale magnetic helicity 
$\langle {\bm a} {\bm \cdot} {\bm b} \rangle(t_0)$ related to the fluctuations of 
the vector potential ${\bm a}$ and the magnetic field 
${\bm b}$ vanishes, we have
$\chi_{\rm tot}(t_0)=0$.
The initial $\mu'_5$ with a wide range of scales produces
${\bm b}$ 
by the chiral dynamo.
Indeed, for a wide spectrum
in $k$ space, fluctuations of $\mu_5$
on larger scales serve as a {\em mean field}
for fluctuations on smaller scales, so that 
the chiral dynamo instability excites ${\bm b}$ and produces
small-scale magnetic helicity
$\langle {\bm a} {\bm \cdot} {\bm b} \rangle$.
Because of the conservation of total chirality, $\chi_{\rm tot}(t)=0$,
the generation of
$\langle {\bm a} {\bm \cdot} {\bm b} \rangle$
causes growth of the mean chiral chemical potential,
$\langle \mu_5 \rangle = - \lambda \langle {\bm a} {\bm \cdot} {\bm b} \rangle /2$.
Simultaneously, 
the chiral dynamo
drives turbulence magnetically and therefore enhances the fluid and magnetic Reynolds numbers,
$\mathrm{Re} \equiv U_\mathrm{rms}/(\nu k_\mathrm{int})$ and 
$\mathrm{Re}_\mathrm{M}\equiv U_\mathrm{rms}/(\eta k_\mathrm{int})$, where $k_\mathrm{int}^{-1}$ 
is the integral scale of magnetically driven turbulence.
When $\mathrm{Re}_\mathrm{M}$ is large enough,
the mean-field dynamo instability 
is excited and amplifies a large-scale magnetic field.
These theoretical ideas are now checked in DNS.

\begin{table}
\small
\centering
\caption{Summary of all runs.}
\begin{tabular}{ l | l l l  l  l }
Run     &  $E_5(k,t_0)$  & $\mu_{5,\mathrm{rms}}(t_0)$ &   $\mu_{5,\mathrm{max}}(t_0)$ &   $\mu_{5,\mathrm{max}}(t_5)$ &  $\mathrm{max}(\Rm)$   \\
 \hline
R$-2$   & $\propto k^{-2}$             & 13.8   & 50.5  & 48.1        & 288  \\
R$-1$   & $\propto k^{-1}$             & 15.8   & 85.8  & 62.0        & 134  \\
R+1     & $\propto k^{1} e^{-(k/10)^2}$   & 12.6   & 53.7  & 53.7        & 65.1  \\
\end{tabular}
\label{tab_DNSoverview}
\end{table}

We use the \textsc{Pencil Code} 
\citep{PC2020} to solve Eqs.~(\ref{ind-DNS})--(\ref{mu-DNS})
with high-order finite difference methods
in a 3D periodic domain of size $L^3 = (2\pi)^3$ with 
a resolution of $672^3$.
The smallest wave number covered in the numerical domain is $k_1 = 2\pi/L = 1$ 
which we use for normalization of length scales. 
All velocities are normalized to $\cs = 1$ 
and the mean fluid density is $\meanrho = 1$.
Time is expressed in terms of the resistive time 
$t_\eta = (\eta k_1^2)^{-1}$ with $\eta$ being the microscopic diffusivity, which is a relevant constant throughout the DNS.
We stress, however, that in magnetically driven turbulence, 
turbulent diffusion dominates shortly after the onset of 
the mean-field dynamo,
yet it is not practical for normalization due to its time dependence.

For numerical stability, diffusion of $\mu_5$ has to be
applied in Eq.~(\ref{mu5-DNS}).
To affect primarily the largest resolved 
wave numbers $k$ in the simulation domain, we use hyperdiffusion,
$\mathscr{D}_5(\mu_5)= -\mathcal{D}_5 \, \nabla^4 \mu_5$;
see the companion paper \citep{SchoberEtAl2022_companion} for technical details.
In all runs, we use $\nu=\eta=2\times10^{-4}$, i.e.\ $\Rm = \mathrm{Re}$, which 
are based on the time-dependent integral scale of magnetically driven turbulence,
\begin{eqnarray}
   k_\mathrm{int}^{-1} \equiv \frac{\int_1^{k_\mathrm{max}}
   E_\mathrm{M}(k)\,k^{-1}~\mathrm{d}k}{\int_1^{k_\mathrm{max}} E_\mathrm{M}(k)~\mathrm{d}k}.
\label{eq_kint}
\end{eqnarray}
Here, $E_\mathrm{M}$ is the magnetic energy spectrum, scaled such that
$\int_0^{k_\mathrm{max}} E_\mathrm{M}(k,t)\,\mathrm{d} k \equiv \bra{\BB^2}/2$.
Likewise, power spectra of $\mu_5$ obey
$\int_0^{k_\mathrm{max}} E_{5}(k,t)\,\mathrm{d} k  \equiv  \bra{\mu_5^2}$.
As initial conditions we use $\UU =\boldsymbol{0}$ and
a weak seed magnetic field in form of Gaussian noise.
Initial fluctuations of $\mu_5$ are
also set up as Gaussian noise, but with a specific
spectrum that follows a power law in $k$ space, i.e.,
$E_5(t_0) = E_{5,0} \left(k/k_1\right)^{s} 
\mathrm{exp}\left(-k^2/k_\mathrm{cut}^2\right)$
with a cutoff $k_\mathrm{cut}$ that is needed for $s>-1$.
We perform runs with $s=-2,-1,+1$ (see Table~\ref{tab_DNSoverview})
and the amplitude $E_{5,0}$ 
is chosen such that the maximum value 
of $\mu_5$ in the domain is comparable
for all runs at the time $t_5$ when the chiral dynamo starts.
In all runs, the initial mean value of $\mu_5$ is vanishing, 
so that $\chi_\mathrm{tot} = \langle \mathcal{H}\rangle + 2 \langle \mu_5 
\rangle/\lambda \approx 0$, and we use $\lambda=400$.

\begin{figure}[t]
\includegraphics[width=0.45\textwidth]{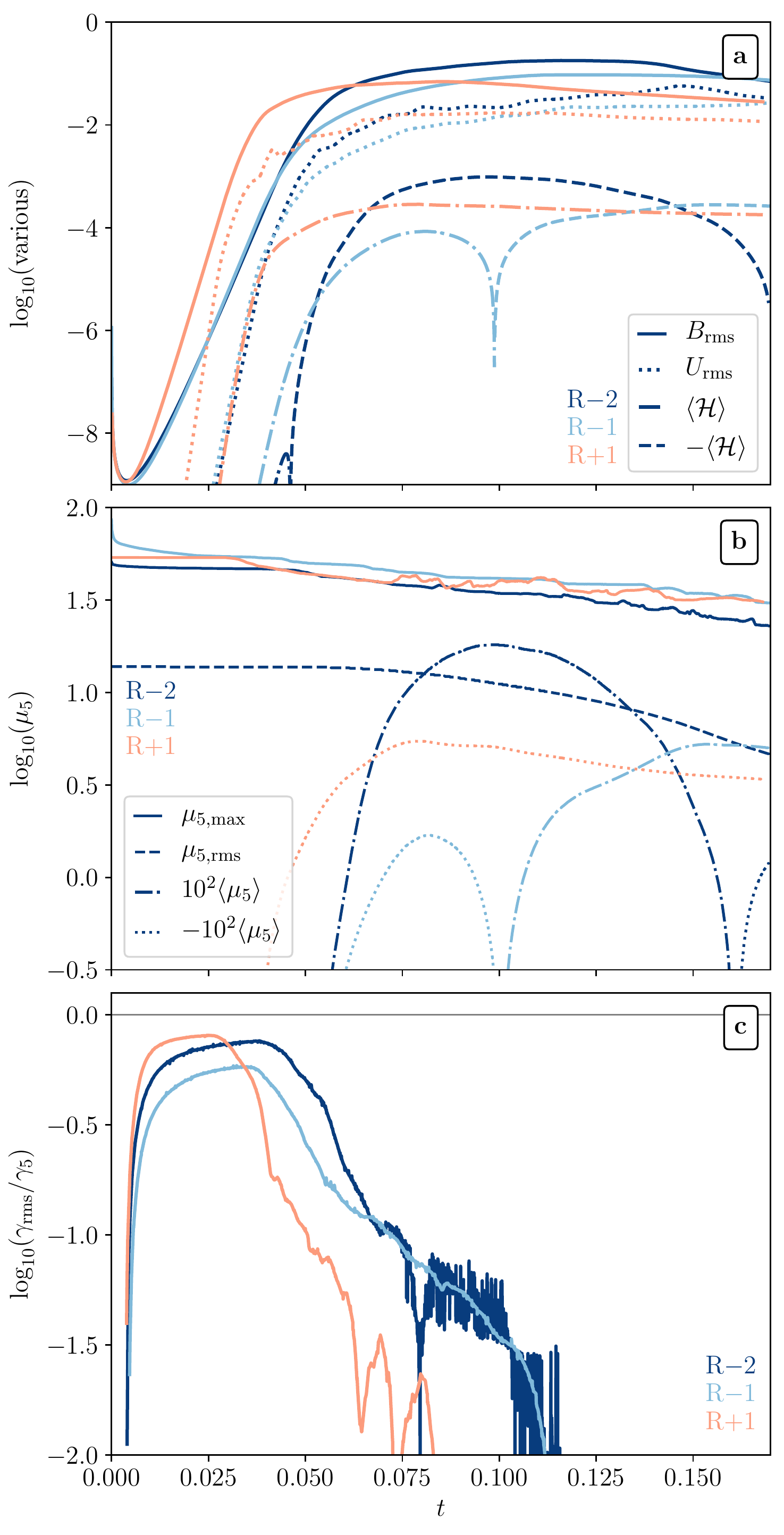} \vspace{-0.4cm}
\caption{
Direct comparison between the time evolution of different 
quantities of all simulations.
Different colors refer to different runs: R$-2$ (dark blue),
R$-1$ (light blue), and R+1 (orange).
\textit{(a)} 
Time series of 
$B_\mathrm{rms}$, $U_\mathrm{rms}$, 
and $\langle \mathcal{H}\rangle$.
\textit{(b)} 
Time series of $\mu_\mathrm{5, max}$ 
and $\langle \mu_5 \rangle$. 
The latter has been multiplied by a factor of $100$ for better visualization. 
\textit{(c)} Measured growth rate of $B_\mathrm{rms}$, $\gamma_\mathrm{rms}$, over
$\gamma_5 = \eta \mu_\mathrm{5, max}^2/4$.
}
\label{fig_t_v5}
\end{figure}

The fluctuations $\mu_5'$ result in an exponential growth of
$B_\mathrm{rms}$ at the rate $\gamma_5$ due to 
the chiral dynamo, as can be seen in Fig.~\ref{fig_t_v5}a.
Usage of $v_5=\eta \mu_{5,\mathrm{max}}$ in the expression for $\gamma_5$ with the maximum value of the chiral 
chemical potential, $\mu_{5,\mathrm{max}}$, as shown in Fig.~\ref{fig_t_v5}b,
reproduces the observed growth rate for all runs rather well;
see Fig.~\ref{fig_t_v5}c (and Fig.~\ref{fig3}b).
We note, however, that a sufficient separation of scales is
required for the dynamo to reach the maximum possible growth
rate; see the accompanying paper \cite{SchoberEtAl2022_companion}.
When comparing the measured growth rate with $\gamma_5$, 
we neglect the change of $\mu_5$ in time, which is much smaller than the increase of $B_\mathrm{rms}$.
During the chiral dynamo
phase, $\langle \mathcal{H}\rangle$
(Fig.~\ref{fig_t_v5}a) and
$\langle \mu_\mathrm{5}\rangle$ [Fig.~\ref{fig_t_v5}b] are produced. 
If the divergence of magnetic helicity fluxes is small, 
the latter two always tend to have opposite signs,
as follows from the conservation of total chirality.
Therefore, contrary to previously considered cases with an initially uniform $\mu_\mathrm{5}$,
the conservation law cannot be used to estimate the maximum magnetic field produced by the 
chiral dynamo.

\begin{figure}
  \includegraphics[width=0.45\textwidth]{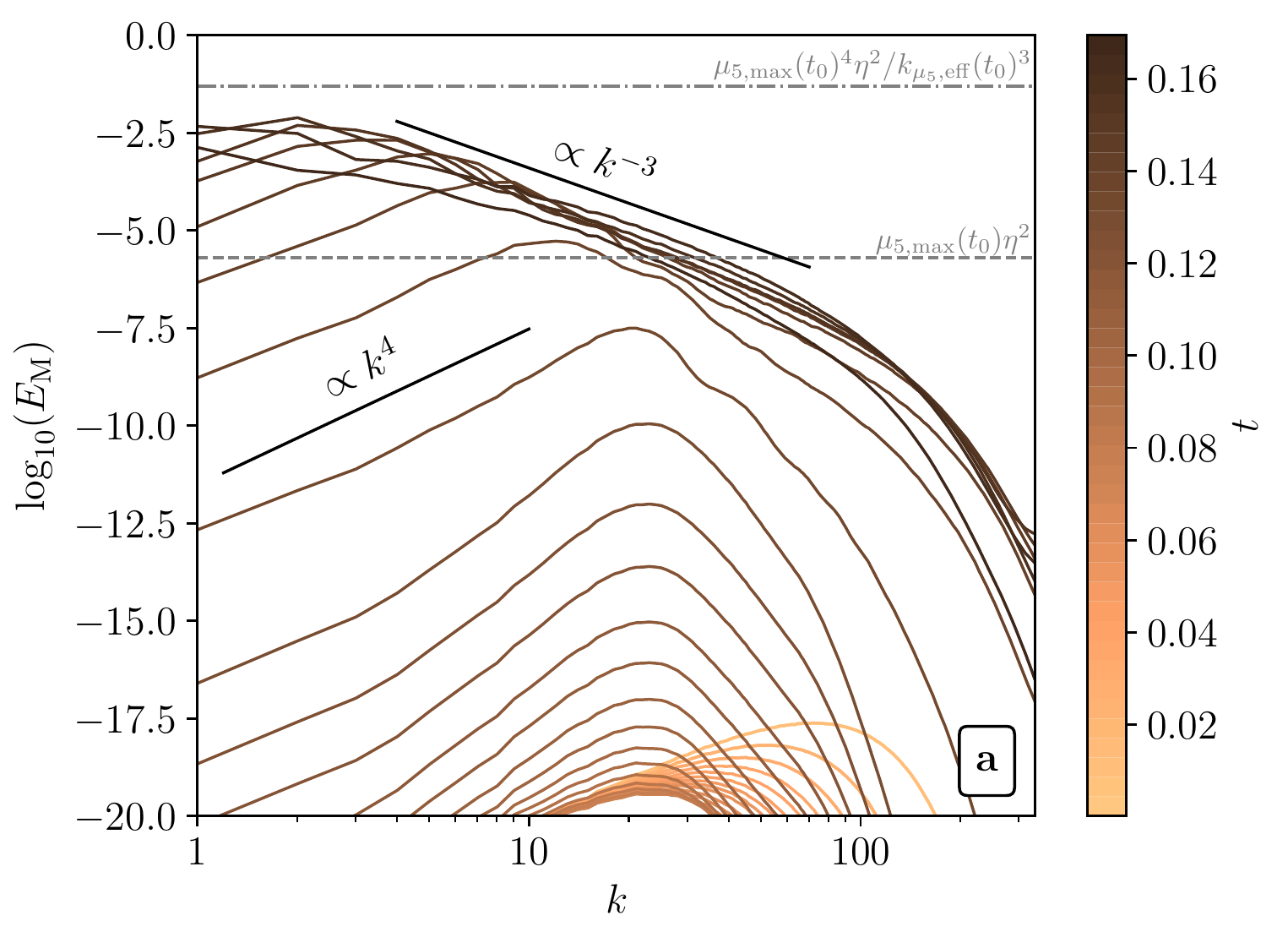}\vspace{-5pt}\\
  \includegraphics[width=0.45\textwidth]{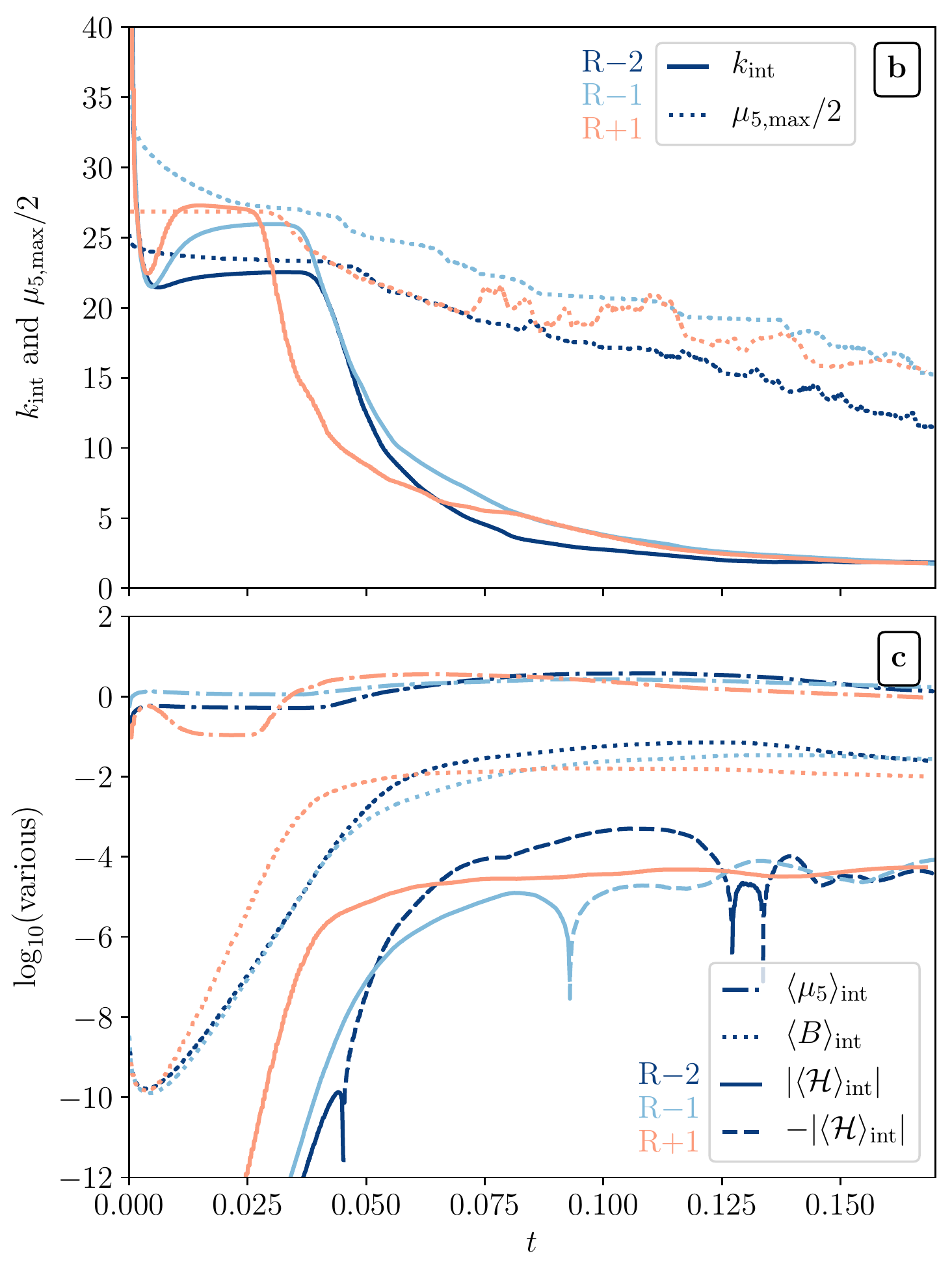} \vspace{-0.4cm}
\caption{
\textit{(a)}
Time evolution of the magnetic energy spectrum $E_\mathrm{M}$ for run R$-2$ with time indicated by the color bar.
\textit{(b)}
The wave number based on the integral scale of turbulence,
$k_\mathrm{int}$, as a function of time for all runs (solid lines)
and the value of the theoretically predicted wave number,
$\mu_{5,\mathrm{max}}/2$, on which the $v_5$ dynamo instability
has the largest growth rate (dotted lines).
\textit{(c)}
Different averages based on the $k_\mathrm{int}$:
$\langle \mu_5 \rangle_\mathrm{int}$ (dashed-dotted), 
$\langle B \rangle_\mathrm{int}$ (dotted lines),  
$\langle\mathcal{H}\rangle_\mathrm{int}$ (solid lines), and 
$-\langle\mathcal{H}\rangle_\mathrm{int}$ (dashed lines).
}
\label{fig_mean}
\end{figure}

With magnetic field amplification via 
the chiral dynamo,
velocity fluctuations are produced by the Lorentz force.
When the turbulent velocity approaches the Alfv\'en speed,
$U_\mathrm{rms} \approx \vA \equiv B_\mathrm{rms}$
(at $t\approx 0.03$ for run R+1 and $t\approx 0.05$ for
runs R$-2$ and R$-1$) the small-scale chiral dynamo phase ends.
This coincides with the time $t_\mathrm{IC}$
when the peak of the magnetic energy spectrum reaches 
$\eta^2 \mu_{5,\mathrm{max}} (t_0)$ and starts to shift 
to larger scales; see $E_\mathrm{M}$ for run R$-2$ in Fig.~\ref{fig_mean}a.

In such chiral-magnetically driven turbulence, 
a mean-field dynamo instability
can occur if
$\mathrm{Re}$ and $\mathrm{Re}_\mathrm{M}$ are large. 
To study the mean-field dynamo, we perform averages
$\langle{\mu_5}\rangle_\mathrm{int}$, $\langle{B}\rangle_\mathrm{int}$,
and $\langle{\mathcal{H}}\rangle_\mathrm{int}$
on the scale $k_\mathrm{int}$ 
(see Fig.~\ref{fig_mean}b), defined as
\begin{eqnarray}
  \langle{X}\rangle_\mathrm{int} &=&
  \left[\frac{\int_0^{k_\mathrm{max}} E_\mathrm{M}(k)\,E_X(k)\,\mathrm{d}k}{\int_0^{k_\mathrm{max}} E_\mathrm{M}(k)~\mathrm{d}k}\right]^{1/2},
\end{eqnarray}
where $E_X(k)$ is the spectrum of $X$; see Fig.\ \ref{fig_mean}c.

The mean-field dynamo instability has a maximum growth rate of 
$\gamma_{\alpha}= (\eta \langle \mu_5 \rangle_\mathrm{int} + \alpha_\mu + \alpha_\mathrm{M} + \alpha_\mathrm{K})^2 /(4\eta_\mathrm{T})$, where 
$\eta_T \approx U_{\rm rms}/(3 k_\mathrm{int})$ is 
the turbulent magnetic diffusivity.
The different $\alpha$ effects are approximately given by
$\alpha_\mu = -(2/3) \eta \langle \mu_5 \rangle_\mathrm{int} \mathrm{log}(\Rm)$ \citep{REtAl17}, 
$\alpha_\mathrm{M} = 2(q-1)/(q+1) \, \tau_{\rm c} \, \chi_{\rm c}$,
and $\alpha_\mathrm{K} = - (1/3) \, \tau_{\rm c} \, \chi_{\rm K}$.
Here, 
$\chi_{\rm c} = \langle {\bm b} {\bm \cdot} ({\bm \nabla} \times{\bm b}) \rangle_\mathrm{int} \approx \langle {\bm a}\cdot {\bm b}\rangle_\mathrm{int} k_\mathrm{int}^2$
is the current helicity,
$\chi_{\rm K} = \langle{{\bm u} {\bm \cdot} {\bm \omega}}\rangle_\mathrm{int}$
is the kinetic helicity, 
${\bm \omega}\equiv \nabla \times {\bm u}$ is the vorticity,
$\tau_{\rm c} \approx (\vA k_\mathrm{int})^{-1}$ 
is the correlation time of magnetically driven turbulence,
and $q$ is the slope of the 
magnetic energy spectrum $\propto k^{-q}$. 
We use $q=3$; see Fig.\ \ref{fig_mean}a.
Figure~\ref{fig3}a shows that $\alpha_\mathrm{M}$ dominates once turbulence
is produced
and therefore the mean-field dynamo growth rate is
$\gamma_{\alpha} \approx \alpha_\mathrm{M}^2/(4\eta_\mathrm{T})$.

\begin{figure}[t]
  \includegraphics[width=0.45\textwidth]{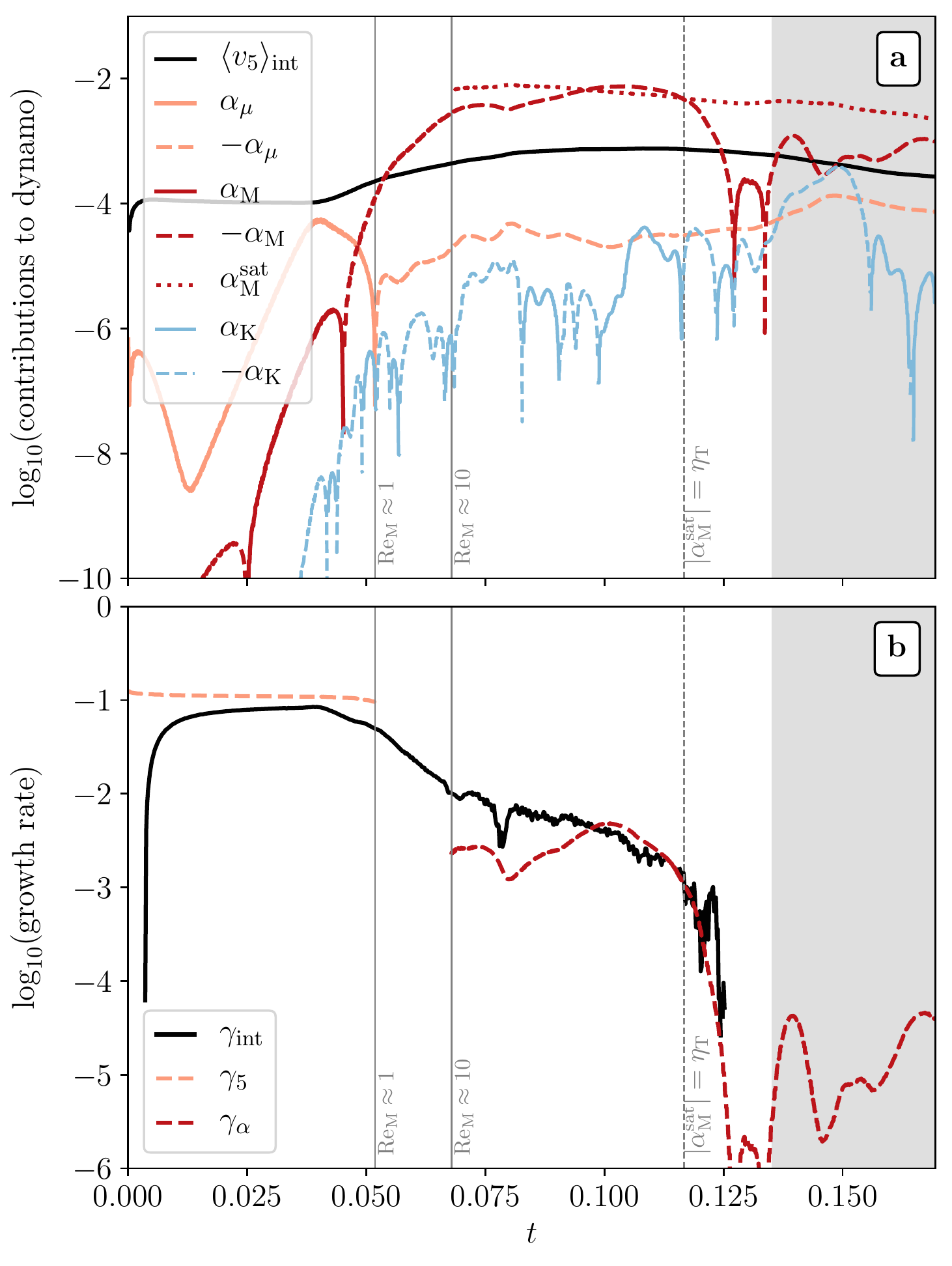}\vspace{-0.4cm}
\caption{
Time evolution of different quantities in Run R$-2$. 
Gray background indicates that 
the inverse cascade has reached the size of the domain.
\textit{(a)}
Different contributions to
the mean-field dynamo growth rate, including
$\langle v_5\rangle_\mathrm{int} \equiv \eta \langle \mu_5\rangle_\mathrm{int}$.
\textit{(b)} 
The measured growth rate of 
$\langle{B}\rangle_\mathrm{int}$, $\gamma_\mathrm{int}$ (black solid line) compared to 
the chiral dynamo
growth rate $\gamma_5$ (orange dashed line) 
and the mean-field dynamo growth rate 
$\gamma_\alpha$ based on $\alpha_\mathrm{M}$
(red dashed line).
}
\label{fig3}
\end{figure}

Our DNS indicate that $\chi_{\rm c}$ plays the key role for the mean-field dynamo sourced by initially
inhomogeneous fluctuations of
$\mu_5$; see Fig.\ \ref{fig3}a and the accompanying paper \cite{SchoberEtAl2022_companion}.
The evolution of $\chi_{\rm c}$ is closely connected to that of the small-scale magnetic helicity \citep{REtAl17}:
\begin{eqnarray}
{\partial \over \partial t}  \overline{{\bm a} {\bm \cdot} {\bm b}} + {\rm div} \, {\bm F}
= 2 \meanv_5 \overline{{\bm b}^2} - 2 \meanEMF \cdot \meanBB
- 2 \eta \, \overline{{\bm b} \, ({\bm \nabla} \times {\bm b})}  ,
\label{MH1}
\end{eqnarray}
where $\meanEMF \equiv \overline{{\bm u} {\bm \times} {\bm b}}=\alpha_\mathrm{M} \meanBB - \eta_T \, ({\bm \nabla} \times \meanBB)$ 
is the electromotive force with $\alpha_\mathrm{M}$ being the 
dominant contribution to the total $\alpha$ effect,
and ${\bm F}$ is the flux of $\overline{{\bm a} {\bm \cdot} {\bm b}}$.
Near magnetic field maximum, two leading source/sink 
terms in Eq.~(\ref{MH1}),
$2 \meanv_5  \overline{{\bm b}^2} - 2 \alpha_\mathrm{M}  \meanBB^2$, 
compensate each other, so that
the magnetic $\alpha$ effect reaches the value
$\alpha_\mathrm{M}^{\rm sat} = \eta \, \meanmu_{5} \, {\overline{{\bm b}^2} / \meanBB^2}$.
For R$-2$, $|\alpha_\mathrm{M}| \approx |\alpha_\mathrm{M}^{\rm sat}|$ for $t\gtrsim0.075$, as can be seen in Fig.~\ref{fig3}a.

\begin{figure}
 \includegraphics[width=0.45\textwidth]{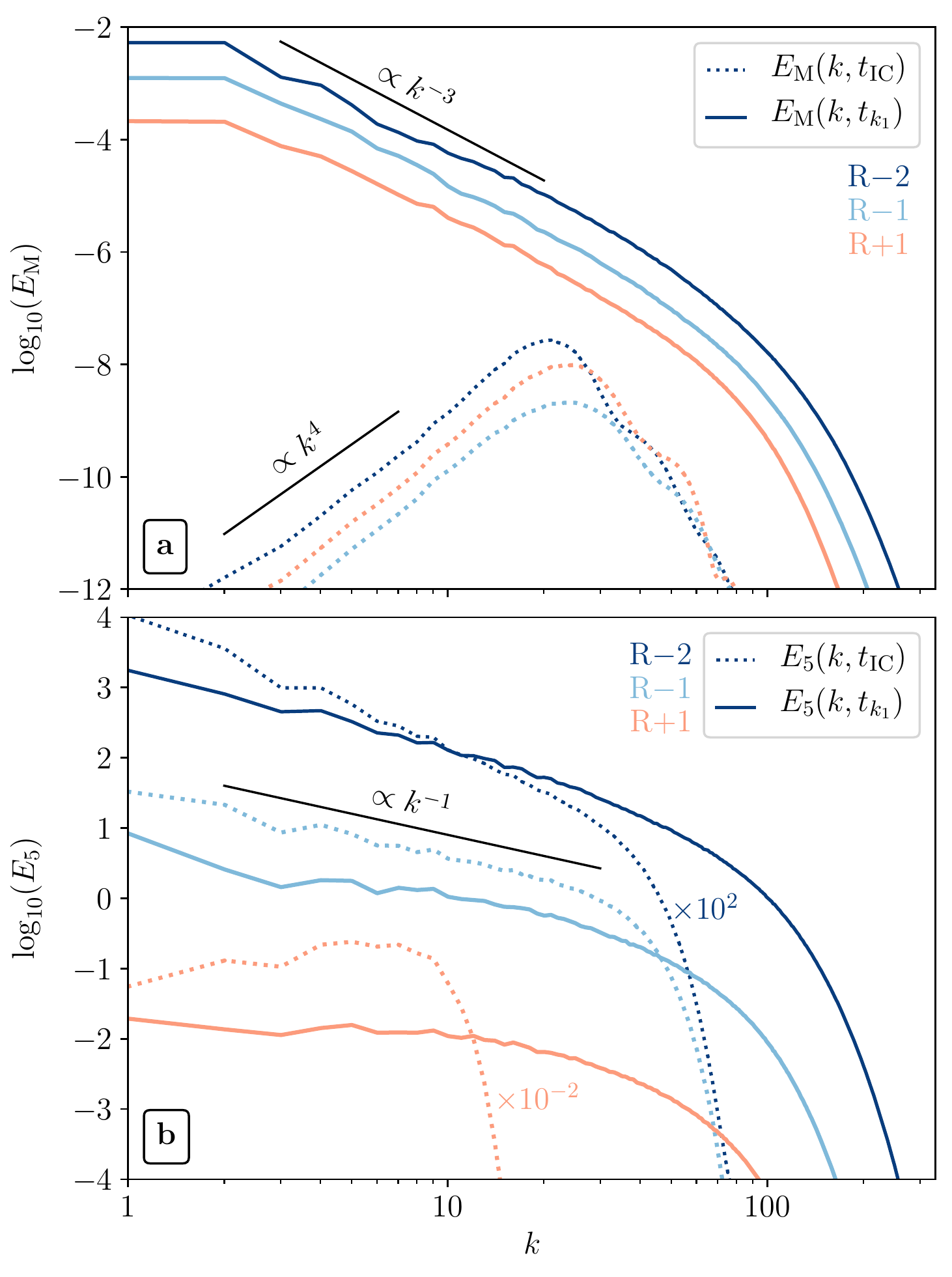}  \vspace{-0.4cm}
\caption{Power spectra from all simulations. 
\textit{(a)} 
Magnetic energy spectra $E_\mathrm{M}$ at the beginning of the chiral 
inverse cascade $t_\mathrm{IC}$ (dotted lines) and the 
time when the cascade reaches the size of the numerical 
domain $t_{k_1}$ (solid lines).
\textit{(b)} 
Spectra of $\mu_5$ shown at the same two characteristic times
as $E_\mathrm{M}$. 
For better visibility the spectra of runs R$-2$ and
R+1 have been multiplied by factors of $10^2$ and $10^{-2}$,
respectively.
}
\label{fig_spectra}
\end{figure}

The maximum growth rate of the mean-field dynamo 
instability $\gamma_{\alpha}$ agrees well with 
the measured growth rate $\gamma_\mathrm{int}$ of $\langle B\rangle_\mathrm{int}$;
see Fig.~\ref{fig3}b for run~R$-2$ in the interval 
$0.075 < t < 0.12$.
Since $\chi_{\rm tot}(t_0)=0$,
the conservation law cannot be employed here
to find the maximum magnetic field value;
but see the companion paper \citep{SchoberEtAl2022_companion} for a phenomenological model.
In our DNS, $\gamma_\mathrm{int}$ strongly decreases when the scale at which $\gamma_{\alpha}$
is maximum becomes larger than the size of the box.
As can be seen in Fig.\ \ref{fig3}b, $\gamma_\mathrm{int}$ 
vanishes once the positive contribution to the growth rate on
the minimum wave number of the box, $|\alpha_\mathrm{M}^\mathrm{sat}| k_1$, 
becomes comparable to the negative contribution, $\eta_\mathrm{T} k_1^2$. 
For R$-2$, dissipation 
due to $\eta_\mathrm{T} k_1^2$
on the box scale dominates for $t\gtrsim0.12$.

At the time $t_{k_1}$ when the peak of the magnetic
energy reaches the size of the domain, 
all of the $\mu_5$ spectra approach a universal $k^{-1}$;
see Fig.\ \ref{fig_spectra}.
The magnetic energy spectra approach a $k^{-3}$ scaling
which is, for fully helical magnetic fields,
consistent with the magnetic helicity spectra $\propto k^{-4}$.

In conclusion, 
a small-scale chiral dynamo can arise from an initially 
fluctuating chiral chemical potential with zero mean. 
The chiral dynamo generates small-scale magnetic helicity which
(i) produces a mean $\mu_5$ due to the conservation of total chirality and 
(ii) drives turbulence via the Lorentz force.
In our DNS, sufficiently strong turbulence is generated to 
activate a mean-field dynamo that is well described by the 
magnetic $\alpha$ effect caused by current helicity.
During the mean-field dynamo phase, the power spectra 
develop a universal shape; 
$E_\mathrm{M} \propto k^{-3}$ and $E_5 \propto k^{-1}$.
In particular, with the onset of turbulence in the system, 
$\mu_5$ becomes scale invariant, independent of its initial condition.

\begin{acknowledgments}
We have benefited from stimulating discussions with
Abhijit B.\ Bendre, Nathan Kleeorin, and Matthias Rheinhardt.
J.S.~acknowledges the support by the Swiss National 
Science Foundation under Grant No.\ 185863.
A.B.~was supported in part through a grant from the Swedish Research Council
(Vetenskapsr{\aa}det, 2019-04234).
\end{acknowledgments}

\end{document}